\documentclass[12pt]{article}
\usepackage{amsfonts}
\usepackage{amsmath}
\usepackage{amssymb}
\usepackage{epsfig}
\usepackage{color}
\usepackage{float}
\usepackage{tabularx}
\usepackage{ulem}
\usepackage{cancel}
\usepackage{braket}

\definecolor{darkblue}{rgb}{0,0,.5}

\begin{document}
\thispagestyle{empty}

\begin{flushright} 
February 2024\\
\end{flushright}
\vspace{6mm}
\begin{center}
\Large {\bf Analytic and Parameter-Free Formula for the Neutrino Mixing Matrix}\\
\mbox{ }\\
\normalsize
\vspace{1.7cm}
{\bf Bodo Lampe} \\              
\vspace{0.4cm}
II. Institut f\"ur theoretische Physik der Universit\"at Hamburg \\
Luruper Chaussee 149, 22761 Hamburg, Germany \\
%e-mail: Lampe.Bodo@web.de \\   

\vspace{1.9cm}

%\begin{figure}[H]
%%%%%[H] in verbindung mit package float
%\begin{center}
%%%%\epsfig{file=fig1philo.eps,height=5.4cm}
%\epsfig{file=Bodolampe2014a.eps,height=6.0cm}
%%%%\epsfig{file=Bodolampe2008.eps,height=6.0cm}
%\end{center}
%\end{figure}

%\vspace{0.8cm}
%\vspace{3.0cm}
{\bf Abstract}
\end{center} 
\vspace{-0.5cm}
A parameter-free analytic expression for the PMNS matrix is derived which fits numerically all the measured matrix components at 99.7$\%$ confidence. Results are proven within the microscopic model and include a prediction of the leptonic Jarlskog invariant. The approach is universal in the sense that it can be applied to the quark sector as well. Preliminary numbers obtained for the CKM matrix elements look promising, but are plagued with large theoretical errors. 

\newpage

\normalsize
%\large
%DIES GIBT GRoeSSERE SCHRIFT

%\section{nicht vergessen} 

%--------------------------------------------------------

%\newpage
%\vspace{3.0cm}

%\vspace{0.3cm}
%\begin{flushright}
%\emph{Der Kampf der Vernunft besteht darin,}\\
%\emph{dasjenige, was der Verstand fixiert hat,}\\ 
%\emph{zu \"uberwinden.}\\
%{\sc G. W. F. Hegel}
%\end{flushright}

%\color{blue}{$\Lambda c^/3$}
%\color{red}{$\Lambda c^/3$}
%\color{green}{$\Lambda c^/3$}

{\bf \underline{Introduction}}

As well known there is a mixing between the flavor and mass eigenstates of the 3 neutrino species, and this can be described by a unitary matrix, the PMNS neutrino mixing matrix\cite{maki,giganti}. The experimentally relevant quantities are the absolute values of the matrix elements, which describe the amount of admixture of the flavor into mass eigenstates, and the leptonic Jarlskog invariant which describes any possible CP violation in the leptonic sector. 

Since the discovery of neutrino oscillations, many models of neutrino mass and mixings have been constructed. The most straightforward approach is to incorporate Dirac neutrino masses into the Standard Model by introducing three right-handed neutrinos coupled to a Higgs field analogously to the quarks and charged leptons. 

Unfortunately, within the SM the values of the mixing parameters cannot be predicted. 

{\bf \underline{Leading symmetric Approximation}}

In a first step a leading order result for the mixing matrix will be derived which is
\begin{eqnarray} 
V_{PMNS}&=&\exp \Biggl\{\frac{i}{{\sqrt{3}}} \begin{bmatrix}
   0 & 1 & 0  \\
    1 & 1 & -1  \\
     0 &  -1 & -1  \\
\end{bmatrix}\Biggr\} \nonumber \\
&=& \begin{bmatrix}
   0.8467-i 0.0300 & -0.1489+i 0.4861 & 0.1532-i 0.00051  \\
    -0.1489-i 0.4861 & 0.5446+i 0.4568 & -0.00433 - i 0.4858  \\
      0.1532-i 0.00051 &  -0.00433 - i 0.4858 & 0.6892-i 0.5153  \\
\end{bmatrix}
\label{pmn0a}
\end{eqnarray}
while an improved formula will be given later in (\ref{p0imp}). 

The leading order expression (\ref{pmn0a}) is a complex, symmetric and unitary matrix, and the absolute values of the matrix elements can be calculated numerically and compared to measurements 
%(the latter including one standard error)
\begin{eqnarray} 
\begin{bmatrix}
   0.843 & 0.510 & 0.153  \\
    0.510 & 0.711 & 0.486  \\
      0.153 &  0.486 & 0.861  \\
\end{bmatrix}
\quad vs. \quad
\begin{bmatrix}
   0.80-0.85 & 0.51-0.58 & 0.142-0.155  \\
    0.23-0.51 & 0.46-0.69 & 0.63-0.78  \\
      0.25-0.53 &  0.47-0.70 & 0.61-0.76  \\
\end{bmatrix}
\label{pmn1a}
\end{eqnarray} 
By inspection one concludes that the agreement is reasonable but not optimal, with the 23 entry being the most critical. 
%(dies sind die aktuellen Zahlen: denn Bei wiki PMNS im Internet steht: As of November 2022, the 3 sigma ranges (99.7% confidence) for the magnitudes of the elements of the matrix. dh Die Zahlen scheinen 3sigma zu sein was wohl 1 standard error entspricht)
%(In statistics, the 68–95–99.7 rule, also known as the empirical rule, is a shorthand used to remember the percentage of values that lie within an interval estimate in a normal distribution.68, 95, and 99.7 prozent of the values lie within one, two, and three standard deviations of the mean, respectively. Andererseits gilt: Since 95 of values fall within two standard deviations of the mean according to the 68-95-99.7 Rule, simply add and subtract two standard deviations from the mean in order to obtain the 95 confidence interval. Notice that with higher confidence levels the confidence interval gets large so there is less precision.)
The first row, which is best measured, is also best fitting. Concerning the other rows, the experimental results in (\ref{pmn1a}) are non-symmetric, though with very large errors. It will be described later, in connection with (\ref{p0imp}) and (\ref{pmnxxa}), how (\ref{pmn0a}) can be improved by additional non-symmetric contributions so that complete agreement within the errors is obtained.

A prediction for the leptonic Jarlskog invariant\cite{jarls} can be calculated from  (\ref{pmn0a}) as  
\begin{eqnarray} 
J_{PMNS}=\Im (V_{e1} V_{\mu 2} \bar V_{e2} \bar V_{\mu 1} ) =-0.0106  
\label{pmn2a}
\end{eqnarray} 
This value is large as compared to the Jarlskog parameter of the CKM matrix\cite{pdg}. $J_{PMNS}$ has not been measured so far, although there are experimental indications that leptonic CP violation is indeed rather large\cite{jarl2}. 

{\bf \underline{Motivation and Proof}}

The model, on which the proof is based\cite{bodohiggs,bodomasses}, starts from a fundamental isospin doublet field $\Psi=(\psi_\uparrow,\psi_\downarrow)$ consisting of two SO(3,1) Dirac fields $\psi_\uparrow$ and $\psi_\downarrow$. Ordinary matter quarks and leptons are considered as excitations of isospin vectors 
% $\langle \vec Q_L\rangle = \langle \vec Q_R\rangle$ where\cite{pcacpaper}
\begin{eqnarray}
\vec Q_L=\frac{1}{4}\Psi^\dagger (1-\gamma_5)\vec \tau\Psi 
\qquad \qquad
\vec Q_R= \frac{1}{4}\Psi^\dagger (1+\gamma_5)\vec \tau\Psi 
\label{eq894}
\end{eqnarray}
of the $\Psi$-field, namely as fluctuations $\delta \vec Q_{L}$ and $\delta \vec Q_{R}$ 
of the ground state values $\langle \vec Q_{L}\rangle$ and $\langle \vec Q_{R}\rangle$.
%\begin{eqnarray}
%\vec Q_{L}=\langle \vec Q_{L}\rangle+\, \delta \vec Q_{L} +\,  O(\delta^2) \qquad \qquad %\vec Q_{R}=\langle \vec Q_{R}\rangle+\, \delta\vec Q_{R} +\,  O(\delta^2)
%\label{eqxxdrt1}
%\end{eqnarray}
$\vec \tau=(\tau_x,\tau_y,\tau_z)$ are the Pauli matrices in `internal' isospin space, whose coordinates will be denoted as $x$, $y$ and $z$.

Note that the corresponding excitations $\delta \Psi$ are fermions, but their dynamics can best be described in terms of isospin vectors (\ref{eq894}). Namely, mass eigenvalues can be calculated using Hamiltonians H which involve interactions of the isospin vectors and then diagonalizing the equations 
\begin{equation} 
\frac{d\vec Q_{L,R}}{dt} = i \, [H, \vec Q_{L,R}] 
\label{txxm32}
\end{equation}
Assuming a suitable tetrahedral configuration for the isospin vectors, 24 eigenvalues arise from (\ref{txxm32}), which are interpreted as the quark and lepton masses\cite{bodohiggs,bodomasses}.  

While the masses correspond to the eigenvalues, CKM and PMNS mixings can be deduced from the eigenvectors. The relation between the eigenvectors, the mass eigenstates and the weak interaction eigenstates are clarified in the following discussion. Thereby, the result (\ref{pmn0a}) and its improvement (\ref{p0imp}) for the PMNS matrix will be obtained. 

The first step is to label the quark and lepton mass states in terms of the vectors $\delta \vec Q$. More in detail, the following definitions are used: 
\begin{eqnarray} 
\ket{\vec S}= \delta \vec Q_L  \qquad\qquad\qquad \ket{\vec T}= \delta \vec Q_R 
\label{nn77}
\end{eqnarray}
Dirac's notation with bra and ket states is applied here to make the mixing relations more transparent. In fact, (\ref{nn77}) are orthonormal vector states and can be used to write down the equations for the neutrino mass eigenstates, as obtained from the diagonalization procedure\cite{bodomasses}
\begin{eqnarray} 
\ket{\nu_{e,m}}&=&\frac{1}{\sqrt{6}} [(\ket{S_x} +\ket{T_x}) +  (\ket{S_y} + \ket{T_y})  + (\ket{S_z} + \ket{T_z}) ] \nonumber \\
\ket{\nu_{\mu, m}}&=&\frac{1}{\sqrt{6}} [ (\ket{S_x} + \ket{T_x}) +\omega (\ket{S_y} + \ket{T_y}) + \bar \omega (\ket{S_z} + \ket{T_z})] \nonumber \\
\ket{\nu_{\tau, m}}&=& \frac{1}{\sqrt{6}} [(\ket{S_x} +\ket{T_x}) +\bar\omega (\ket{S_y} + \ket{T_y}) + \omega  (\ket{S_z} + \ket{T_z}) ] 
\label{allnn5}
\end{eqnarray}
The corresponding result for the charged leptons is
\begin{eqnarray} 
\ket{e_m}&=&\frac{1}{\sqrt{6}} [  (\ket{T_x}-\ket{S_x}) +  (\ket{T_y}-\ket{S_y})  + (\ket{T_z}-\ket{S_z}) ] \nonumber \\
\ket{\mu_m}&=& \frac{1}{\sqrt{6}}[(\ket{T_x}-\ket{S_x}) + \omega (\ket{T_y}-\ket{S_y})  + \bar \omega (\ket{T_z}-\ket{S_z}) ] \nonumber  \\
\ket{\tau_m}&=& \frac{1}{\sqrt{6}}[(\ket{T_x}-\ket{S_x}) + \bar \omega (\ket{T_y}-\ket{S_y})  +\omega (\ket{T_z}-\ket{S_z}) ] 
\label{allee5}
\end{eqnarray}

The appearance of the complex numbers
\begin{eqnarray} 
\omega=-\frac{1-i\sqrt{3}}{2} \qquad\qquad\qquad \bar\omega=-\frac{1+i\sqrt{3}}{2}
\label{mack151}
\end{eqnarray}
corresponding to rotations by 120 and 240 degrees are an effect of the underlying tetrahedral symmetry. They turn the expressions (\ref{allnn5}) and (\ref{allee5}) into symmetry adapted functions.

The lepton mass states actually can be brought to the much more compact form 
\begin{eqnarray} 
\begin{bmatrix}  \ket{\nu_{e m}}\\     \ket{\nu_{\mu m} }   \\  \ket{\nu_{\tau m} }    \\ \end{bmatrix}
=Z\begin{bmatrix}  \ket{V_x}\\     \ket{V_y}   \\     \ket{V_z}    \\ \end{bmatrix}
\qquad\qquad\qquad\qquad
\begin{bmatrix}  \ket{e_m}\\     \ket{\mu_m}   \\     \ket{\tau_m}    \\ \end{bmatrix}
=Z\begin{bmatrix}  \ket{A_x}\\     \ket{A_y}   \\     \ket{A_z}    \\ \end{bmatrix}
\label{fo77}
\end{eqnarray}
by using the quantities 
\begin{eqnarray} 
\ket{\vec V}= \frac{1}{\sqrt{2}}(\ket{\vec S}+\ket{\vec T})  \qquad \qquad\qquad
\ket{\vec A}= \frac{1}{\sqrt{2}}(\ket{\vec T}-\ket{\vec S}) 
\label{nn77a}
\end{eqnarray}
and the $Z_3$ Fourier transform matrices
\begin{eqnarray} 
Z=\frac{1}{\sqrt{3}}
\begin{bmatrix}
   1 & 1 & 1  \\
    1 & \omega & \bar\omega   \\
    1 &  \bar\omega &  \omega   \\
\end{bmatrix}
\qquad \qquad 
Z^\dagger=\frac{1}{\sqrt{3}}
\begin{bmatrix}
   1 & 1 & 1  \\
  1 &  \bar\omega & \omega   \\
   1 &   \omega &  \bar\omega   \\
\end{bmatrix}
\label{matz35}
\end{eqnarray}
It is interesting to note that the eigenfunctions (\ref{allnn5}), (\ref{allee5}) and (\ref{fo77}) are stable against variations of all the isospin couplings one may use in the Hamiltonian H in (\ref{txxm32}). In consequence, the neutrino mixing matrix does not depend on any fermion mass values. This implies a stable and unambiguous prediction for the PMNS matrix and is in contrast to the CKM matrix in the quark sector, where a mass dependence shows up, cf. Eq. (\ref{wck41111}) later. 

As well known, the defining equation for the PMNS matrix is
\begin{eqnarray} 
%-\frac{g}{\sqrt{2}} 
\begin{bmatrix}  \bra{ \nu_{ew}} & \bra{ \nu_{\mu w}} & \bra{ \nu_{\tau w}} \\ \end{bmatrix} W_\mu^+ 
\begin{bmatrix}  \ket{e_w} \\ \ket{\mu_w}  \\ \ket{\tau_w} \\ \end{bmatrix}
=\begin{bmatrix}  \bra{ \nu_{em}} & \bra{ \nu_{\mu m}} & \bra{ \nu_{\tau m}} \\ \end{bmatrix} W_\mu^+ V_{PMNS}
\begin{bmatrix}  \ket{e_m} \\ \ket{\mu_m}  \\ \ket{\tau_m} \\ \end{bmatrix}
\label{wbwwc}
\end{eqnarray} 
where the index $w$ denotes weak interaction eigenstates, and it is understood that we talk about left handed fields only. The mixing matrix is formally given by
\begin{eqnarray} 
V_{PMNS}=V_N V_L^\dagger = 
\begin{bmatrix}
V_{1e} & V_{1\mu} & V_{1\tau} \\
V_{2e} & V_{2\mu} & V_{2\tau} \\
V_{3e} & V_{3\mu} & V_{3\tau}
\end{bmatrix} 
\label{wb111c}
\end{eqnarray} 
where
\begin{eqnarray} 
V_N=
\begin{bmatrix}
\bra{\nu_{e m}}  \\ \bra{\nu_{\mu m}}\\ \bra{\nu_{\tau m}} \\ 
\end{bmatrix}
\begin{bmatrix}
\ket{\nu_{e w}}  & \ket{\nu_{\mu w}}& \ket{\nu_{\tau w}} \\ 
\end{bmatrix}
\qquad
V_L^\dagger=
\begin{bmatrix}
\bra{e_w}  \\ \bra{\mu_w} \\ \bra{\tau_w} \\ 
\end{bmatrix}
\begin{bmatrix}
\ket{e_m}  & \ket{\mu_m}& \ket{\tau_m} \\ 
\end{bmatrix}
\label{wb222c}
\end{eqnarray} 
Replacing the mass eigenstates by the isospin excitations according to (\ref{fo77}) one obtains
\begin{eqnarray} 
V_{PMNS}=Z 
\Biggl\{
\begin{bmatrix}
\bra{V_x}  \\ \bra{V_y} \\ \bra{V_z} \\ 
\end{bmatrix}
%%%%%%%%%%%\biggl\{
\begin{bmatrix}  \ket{ \nu_{ew}} & \ket{ \nu_{\mu w}} & \ket{ \nu_{\tau w}} \\ \end{bmatrix}
\begin{bmatrix}
\bra{e_w}  \\ \bra{\mu_w} \\ \bra{\tau_w} \\ 
\end{bmatrix}
%%%%%%%%%%\biggr\}
\begin{bmatrix}
\ket{A_x}  & \ket{A_y}& \ket{A_z} \\ 
\end{bmatrix}
\Biggr\} Z^\dagger
\label{wb333c}
\end{eqnarray} 
By inspection one sees that (\ref{wb333c}) exactly compensates all the matrix transformations in (\ref{wbwwc}) and (\ref{fo77}) so as to maintain lepton universality and keep the weak current diagonal in the weak eigenstates.

The brace in (\ref{wb333c}) comprises a matrix of expectation values of the form \begin{eqnarray} 
Y:=\begin{bmatrix} \bra{V_x}  \\ \bra{V_y} \\ \bra{V_z} \\ \end{bmatrix}
\mathcal{O}
\begin{bmatrix}
\ket{A_x}  & \ket{A_y}& \ket{A_z} \\ 
\end{bmatrix}
\label{wb8c}
\end{eqnarray} 
where the inner product 
\begin{eqnarray} 
\mathcal{O}:= 
\begin{bmatrix}  \ket{ \nu_{ew}} & \ket{ \nu_{\mu w}} & \ket{ \nu_{\tau w}} \\ \end{bmatrix}
\begin{bmatrix}
\bra{e_w}  \\ \bra{\mu_w} \\ \bra{\tau_w} \\ 
\end{bmatrix}
\label{dyop1}
\end{eqnarray} 
is a dyadic 1-dimensional operator which acts between the complex 3-dimensional spaces of charged lepton ($\sim \vec S -\vec T$) and antineutrino ($\sim \vec S +\vec T$) states. One may say that it contains all information about what the charged W-boson does to the lepton fields: it changes isospin, mixes families and so on. Weak SU(2) and tetrahedral symmetry force $\mathcal{O}$ to have the form
\begin{eqnarray} 
\mathcal{O}&=&\ket { S_x} \bra  { T_x} + \ket { S_y} \bra  { T_y} + \ket { S_z} \bra  { T_z} -  \ket { T_x} \bra  { S_x} - \ket { T_y} \bra  { S_y} - \ket { T_z} \bra  { S_z}\nonumber \\
&& +\frac{i}{\sqrt{3}}[\ket { S_y} \bra  { S_z} + \ket { S_z} \bra  { S_y} - \ket { T_y} \bra  { T_z} - \ket { T_z} \bra  { T_y}]\nonumber \\
&& +\frac{i}{\sqrt{3}}[\omega \ket { S_x} \bra  { S_y} + \bar \omega \ket { S_y} \bra  { S_x} - \omega \ket { T_x} \bra  { T_y} -  \bar \omega \ket { T_y} \bra  { T_x}]\nonumber \\
&& +\frac{i}{\sqrt{3}}[ \bar \omega \ket { S_x} \bra  { S_z} + \omega \ket { S_z} \bra  { S_x} -  \bar \omega \ket { T_x} \bra  { T_z} - \omega \ket { T_z} \bra  { T_x}]  
\label{wb1}
\end{eqnarray} 
In order to derive (\ref{wb1}) one has to note that SU(2) invariance allows the appearance of dot products and triple products only. The coefficients of these products are then dictated by the tetrahedral symmetry of the isospin vectors. For example, to derive the triple product coefficients one should remember that the $W^+$-boson is defined in the 3 internal dimensions in an analogous manner as a plus circularly polarized wave in 3 spatial dimensions, namely by means of an (internal) `polarization vector' $\vec e_+=(\vec e_1 + i \vec e_2)/\sqrt{2}$ which is perpendicular to the axis of quantization, in this case given by $\sim (1,1,1)$.
\begin{eqnarray} 
\vec e_1 = \frac{1}{\sqrt{2}} (0,1,-1) \qquad  \qquad \vec e_2=\frac{1}{\sqrt{6}} (-2,1,1)
\label{pol11}
\end{eqnarray} 
Introducing the vector
\begin{eqnarray} 
\vec \Omega = \frac{1}{\sqrt{3}} (1,\omega,\bar \omega) 
\label{pol11999}
\end{eqnarray} 
allowed contributions to $\mathcal{O}$ are of the triple product form
\begin{eqnarray} 
&& \varepsilon_{ijk}  \frac{1}{\sqrt{2}}(\vec e_1+i \vec e_2)_i \ket { Q_j} \bra  { Q'_k}
= -\frac{i}{\sqrt{3}} \, \vec \Omega (\vec Q \times \vec Q') 
= -\frac{i}{\sqrt{3}} \, [\,\ket { Q'_y} \bra  { Q_z} - \ket { Q'_z} \bra  { Q_y}\nonumber \\
&& \qquad\qquad -\omega (\ket { Q'_x} \bra  { Q_z} - \ket { Q'_z} \bra  { Q_x})
+ \bar \omega (\ket { Q'_x} \bra  { Q_y} - \ket { Q'_y} \bra  { Q_x}\,) \,]
\label{pol12}
\end{eqnarray} 
for the ket and bra states belonging to any 2 internal angular momenta $Q$ and $Q'$. These contributions are anti-hermitian, and care must be taken in the definition of the complex triple product when using complex conjugation in the determination of $\mathcal{O}$. 

Note that $\mathcal{O}$ as given in (\ref{wb1}) is universal in the sense that it depends only on properties of the $\Psi$ field, and therefore will appear in identical form within the quark sector and the calculation of the CKM matrix. This fact reflects the quark lepton universality of the W-boson interactions.

Inserting (\ref{wb1}) into (\ref{wb8c}) one obtains 
\begin{eqnarray} 
Y=\begin{bmatrix} \bra{V_x}  \\ \bra{V_y} \\ \bra{V_z} \\ \end{bmatrix}
\mathcal{O}
\begin{bmatrix}
\ket{A_x}  & \ket{A_y}& \ket{A_z} \\ 
\end{bmatrix}
=I+X
\label{wb8cc}
\end{eqnarray} 
i.e. a sum of a hermitian part (the unit matrix $I$) and an anti-hermitian matrix
\begin{eqnarray} 
X=-\frac{i}{\sqrt{3}}\begin{bmatrix}
   0 & \bar\omega & \omega  \\
    \omega & 0 & 1  \\
     \bar \omega &  1 & 0  \\
\end{bmatrix} 
\label{matz325}
\end{eqnarray}
The invariant structure which gives the unit matrix in (\ref{wb8cc}) is the dot product, while the invariant structure belonging to the anti-hermitian contribution X is the triple product. The unit matrix corresponds to no mixing at all, so the origin of a non-trivial PMNS matrix is to be found solely in the triple product terms (\ref{pol12}). 

Since the result (\ref{wb8cc}) is not unitary but anti-hermitian, an exponentiation suggests itself which gives a unitary PMNS matrix of the form
\begin{eqnarray} 
V_{PMNS}&=&Z e^{X} Z^\dagger =e^{Z X Z^\dagger}  \nonumber \\
&=& \frac{1}{3}\begin{bmatrix}
   1 & 1 & 1  \\
    1 & \omega & \bar\omega   \\
    1 &  \bar\omega &  \omega   \\
    \end{bmatrix}
%e^{\frac{i}{{\sqrt{3}}} \begin{bmatrix}
%   0 & \omega & \bar\omega  \\
%    \bar\omega & 0 & 1  \\
%      \omega &  1 & 0  \\
%\end{bmatrix}}
\exp \Biggl\{\frac{-i}{{\sqrt{3}}} \begin{bmatrix}
   0 & \bar\omega & \omega  \\
    \omega & 0 & 1  \\
     \bar \omega &  1 & 0  \\
\end{bmatrix}\Biggr\}
\begin{bmatrix}
   1 & 1 & 1  \\
  1 &  \bar\omega & \omega   \\
   1 &   \omega &  \bar\omega   \\
\end{bmatrix}  \nonumber \\
&=& \begin{bmatrix}
   0.8467-i 0.0300 & -0.1489+i 0.4861 & 0.1532-i 0.00051  \\
    -0.1489-i 0.4861 & 0.5446+i 0.4568 & -0.00433 - i 0.4858  \\
      0.1532-i 0.00051 &  -0.00433 - i 0.4858 & 0.6892-i 0.5153  \\
\end{bmatrix}
\label{pmn00a}
\end{eqnarray}
identical to what was claimed in (\ref{pmn0a}).

%folgendes rausgenommen weil die Zahlen für die 2. ordnung nicht stimmen, weil die Formel pmtrub für die 4. Ordnung rauskommt
%One should mention that expansion of the exponential in (\ref{pmn0a}) is convergent, and that truncation of the series at second order already gives a good approximation  
%\begin{eqnarray} 
%V_{PMNS}\approx 1+ Z X Z^\dagger + \frac{1}{2} (Z X Z^\dagger)^2 = \begin{bmatrix}
% \frac{5}{6} & -\frac{1}{6}+\frac{i}{\sqrt{3}} & \frac{1}{6}  \\
%-\frac{1}{6}+\frac{i}{\sqrt{3}}   & \frac{1}{2}+\frac{i}{\sqrt{3}}  & -\frac{i}{\sqrt{3}}  \\
%      \frac{1}{6} &  -\frac{i}{\sqrt{3}} & \frac{2}{3}-\frac{i}{\sqrt{3}}  \\
%\end{bmatrix}
%\label{pmtrua}
%\end{eqnarray} 
%with absolute values 
%\begin{eqnarray} 
%|V_{PMNS}|\approx 
%\begin{bmatrix}
%   0.83 & 0.51 & 0.16  \\
%    0.51 & 0.69 & 0.48  \\
%    0.17 &  0.48 & 0.84  \\
%\end{bmatrix}
%\label{pmtrub}
%\end{eqnarray} 
%Actually, this is the order to which (\ref{pmn0a}) is proven rigorously, assuming unitarity together with the arguments in connection with (\ref{pol12}). Comparing (\ref{pmtrub}) to (\ref{pmn1a}), the approximation seems good for the absolute values. There is no trustworthy outcome, however, for the Jarlskog invariant, as can be seen by working out (\ref{pmn2a}) for  the matrix (\ref{pmtrua}). 

{\bf \underline{Improved Formula for the PMNS Matrix}}

So far only dot product and triple product terms (\ref{pol12}) have been considered as contributing to the operator (\ref{wb1}) and the PMNS result. Actually, there is a third kind of term that needs consideration. Using $\vec\Omega^2 =0$ and the same normalization as in (\ref{pol12}) it is of the form 
\begin{eqnarray} 
-(\vec \Omega \times \vec Q) \, (\vec \Omega \times \vec Q')= (\vec \Omega \vec Q) \, (\vec \Omega \vec Q')
\label{w8899}
\end{eqnarray} 
In the microscopic model, quark and lepton masses are related to torsional, Heisenberg and Dzyaloshinskii isospin interactions of the fundamental $\Psi$ field. Furthermore, as shown in \cite{bodoprep}, these three types of interactions completely fix the structure of the model. 

This fact is reflected in the contributions to the operator $\mathcal{O}$: while the dot products and triple products appearing in (\ref{wb1}) parallel the torsional and Heisenberg interactions, (\ref{w8899}) corresponds to the Dzyaloshinskii Hamiltonian. Working out the products $\ket {Q_i} \bra  {Q'_j}$ arising from (\ref{w8899}), it leads to an additional contribution to (\ref{wb1}) which can be comprised by a matrix
\begin{eqnarray} 
D:= \frac{1}{3}
 \begin{bmatrix}
   1 & \omega & \bar \omega  \\
    \omega & \bar \omega & 1  \\
     \bar \omega & 1 & \omega  \\
\end{bmatrix}
\label{ddma1}
\end{eqnarray}
The role of D for (\ref{w8899}) is analogous to that of X for the triple product term.
Combining the X and D contributions an improved formula for the PMNS matrix is obtained
\begin{eqnarray} 
V_{PMNS}=
\exp \Biggl\{  \frac{1}{3}\begin{bmatrix}
   0 & 0 & 0  \\
    0 & 0 & 1  \\
     0 &  -1 & 0  \\
\end{bmatrix}\Biggr\} 
\,
\exp \Biggl\{\frac{i}{{\sqrt{3}}} \begin{bmatrix}
   0 & 1 & 0  \\
    1 & 1 & -1  \\
     0 &  -1 & -1  \\
\end{bmatrix}\Biggr\} 
\label{p0imp}
\end{eqnarray}
This represents a complex and unitary matrix whose absolute value matrix $|V_{PMNS}|$ is not symmetric, in contrast to (\ref{pmn0a}). Its elements are given by
\begin{eqnarray} 
\begin{bmatrix}
   0.847 & 0.510 & 0.153  \\
    0.468 & 0.581 & 0.666  \\
      0.251 &  0.635 & 0.730  \\
\end{bmatrix}
\quad vs. \quad
\begin{bmatrix}
   0.80-0.85 & 0.51-0.58 & 0.142-0.155  \\
    0.23-0.51 & 0.46-0.69 & 0.63-0.78  \\
      0.25-0.53 &  0.47-0.70 & 0.61-0.76  \\
\end{bmatrix}
\label{pmnxxa}
\end{eqnarray} 
and fit the phenomenological numbers to within one standard error. 

The value of the leptonic Jarlskog invariant now is 
\begin{eqnarray} 
J_{PMNS}=0.0454
\label{pnw33}
\end{eqnarray} 
Thus, while the improvement (\ref{p0imp}) only moderately corrects  the absolute values, it strongly modifies the prediction for $J_{PMNS}$. This is because - in contrast to the absolute values - the Jarlskog invariant is dominated by higher orders of the exponential expansion.  

{\bf \underline{Application to the Quark Sector}}

Mixing in the quark sector has been known since the time of Cabibbo\cite{cab}. Although the mixing percentages are smaller, it is much better measured than in the lepton sector. On the other hand, concerning theory, the predictions for the CKM mixing elements  in the present model are somewhat more difficult to obtain, though parts of the arguments for leptons can be taken over to the quark sector. The idea is again that the mixing matrix counterbalances the deviation of the mass eigenstates from the weak eigenstates in such a way that the charged current effectively acts diagonal on the isospin operators (\ref{nn77}). The main complication is the appearance of mass dependent factors in the quark eigenstates, see below. 

The CKM matrix is defined analogously to the PMNS matrix (\ref{wb111c}) and (\ref{wb222c}) 
 %Concentrating on the quarks for the moment, in the SM the mass eigenstates are obtained by diagonalizing the Yukawa matrices $Y_U$ and $Y_D$ of up- and down-type quarks by 4 unitary matrices $V_{LU}$, $V_{RU}$, $V_{LD}$  and  $V_{RD}$ according to 
%\begin{eqnarray} 
%M_U=v\, V_{LU} Y_U V_{RU}^\dagger \qquad\qquad\qquad M_D=v\, V_{LD} Y_D V_{RD}^\dagger 
%\label{mack11}
%\end{eqnarray}
%where $v$ is the vacuum expectation value of the Higgs field. The SM charged current of the quarks is then given by 
%\begin{eqnarray} 
%j_\mu^+=g\, [\bar u_L,\bar c_L, \bar t_L] \gamma_\mu V_{CKM} 
%\begin{bmatrix}
%d_L \\
%s_L \\
%b_L \\
%\end{bmatrix} +c.c.
%\label{mack12}
%\end{eqnarray}
\begin{eqnarray}
V_{CKM}&=&V_{U} V_{D}^\dagger = 
\begin{bmatrix}
V_{ud} & V_{us} & V_{ub} \\
V_{cd} & V_{cs} & V_{cb} \\
V_{td} & V_{ts} & V_{tb}
\end{bmatrix} \nonumber \\
&=& \begin{bmatrix}
\langle u_m | u_w \rangle & \langle u_m | c_w \rangle & \langle u_m | t_w \rangle\\
\langle c_m | u_w \rangle & \langle c_m | c_w \rangle & \langle c_m | t_w \rangle\\
\langle t_m | u_w \rangle & \langle t_m | c_w \rangle & \langle t_m | t_w \rangle
\end{bmatrix}
\begin{bmatrix}
\langle d_w | d_m \rangle & \langle d_w | s_m \rangle & \langle d_w | b_m \rangle\\
\langle s_w | d_m \rangle & \langle s_w | s_m \rangle & \langle s_w | b_m \rangle\\
\langle b_w | d_m \rangle & \langle b_w | s_m \rangle & \langle b_w | b_m \rangle
\end{bmatrix}
\label{mmccm}
\end{eqnarray}
where $m$ denotes mass eigenstates (the physical states) and $w$ weak interaction eigenstates. 

Solving the eigenvalue problem (\ref{txxm32}) leads to mass eigenstates for the up-type quarks
\begin{eqnarray} 
u_m&=&\frac{1}{\sqrt{3}\sqrt{1+\epsilon_{1}^2}}  [(\ket{S_x} +\epsilon_{1} \ket{T_x}) +  (\ket{S_y} + \epsilon_{1} \ket{T_y})  + (\ket{S_z} + \epsilon_{1} \ket{T_z}) ] 
%= R Z \begin{bmatrix} \bra{S_x}  \\ \bra{S_y} \\ \bra{S_z} \\ \end{bmatrix} +R E Z
%\begin{bmatrix} \bra{T_x}  \\ \bra{T_y} \\ \bra{T_z} \\ \end{bmatrix}
\nonumber \\
c_m&=& \frac{1}{\sqrt{3}\sqrt{1+\epsilon_{2}^2}} [ (\ket{S_x} +\epsilon_2 \ket{T_x}) +\omega (\ket{S_y} + \epsilon_2 \ket{T_y}) + \bar \omega (\ket{S_z} + \epsilon_2 \ket{T_z})] \nonumber \\
t_m&=& \frac{1}{\sqrt{3}\sqrt{1+\epsilon_{3}^2}} [(\ket{S_x} +\epsilon_{3} \ket{T_x}) +\bar\omega (\ket{S_y} + \epsilon_{3} \ket{T_y}) + \omega  (\ket{S_z} + \epsilon_{3} \ket{T_z}) ]
\label{allup}
\end{eqnarray}
and for the down quarks
\begin{eqnarray} 
d_m&=&\frac{1}{\sqrt{3}\sqrt{1+\epsilon_{1}^2}} [  (\ket{T_x}-\epsilon_1 \ket{S_x}) +  (\ket{T_y}-\epsilon_1 \ket{S_y})  + (\ket{T_z}-\epsilon_1 \ket{S_z}) ] \nonumber \\
s_m&=& \frac{1}{\sqrt{3}\sqrt{1+\epsilon_{2}^2}} [(\ket{T_x}-\epsilon_2 \ket{S_x}) + \omega (\ket{T_y}-\epsilon_2 \ket{S_y})  + \bar \omega (\ket{T_z}-\epsilon_2 \ket{S_z}) ] \nonumber  \\
b_m&=& \frac{1}{\sqrt{3}\sqrt{1+\epsilon_{3}^2}} [(\ket{T_x}-\epsilon_3 \ket{S_x}) + \bar \omega (\ket{T_y}-\epsilon_3 \ket{S_y})  +\omega (\ket{T_z}-\epsilon_3 \ket{S_z}) ]
\label{alldown}
\end{eqnarray}
Three coefficients $\epsilon_{1,2,3}$ appear in these equations, which depend on the quark and even on the lepton masses. They can be calculated within the model. Namely, as proven in \cite{bodoprep}, each $\epsilon_{i}$ to a very good approximation only depends on the quark and charged lepton masses of the i-th family. More precisely, one can derive 
the formula\cite{bodoprep}
\begin{eqnarray} 
\epsilon_{i} = \frac{1}{6} \frac{M_{Li}}{M_{Ui}+M_{Di}}
\label{sm14exxx}
\end{eqnarray}
where $M_{Ui}$, $M_{Di}$ and $M_{Li}$ denote the corresponding masses within family i.

By inspection one sees that the lepton eigenfunctions (\ref{allnn5}) and (\ref{allee5}) are recovered from  (\ref{allup}) and (\ref{alldown}) by chosing $\epsilon_{3} = \epsilon_{2} = \epsilon_{1} = 1$. It should be stressed, however, that this is only formally true, because the quark states are defined in a different space than the lepton states. The point is that for simplicity reference has been made so far to only one of the four isospins I, II, III and IV on the tetrahedral structure. While the contributions from I-IV to the lepton states are identical and of the form I+II+III+IV, the generic form of the quark states turns out to be 3$\times$I-II-III-IV, 3$\times$II-I-III-IV and 3$\times$III-II-IV for the 3 colors, respectively.  

Knowing the eigenstates (\ref{allup}) and (\ref{alldown}) one may write down the CKM matrix in an analogous fashion as the PMNS matrix (\ref{wb333c}) for leptons
\begin{eqnarray} 
V_{CKM}=\biggl\{ R Z  
\begin{bmatrix} \bra{S_x}  \\ \bra{S_y} \\ \bra{S_z} \\ \end{bmatrix}
+REZ \begin{bmatrix} \bra{T_x}  \\ \bra{T_y} \\ \bra{T_z} \\ \end{bmatrix} 
\biggr\}
\begin{bmatrix} \ket{ u_w}  & \ket{ c_w}& \ket{ t_w} \\ \end{bmatrix}
\begin{bmatrix} \bra{d_w}  \\ \bra{s_w} \\ \bra{b_w} \\ \end{bmatrix}
\times \nonumber\\
\times \biggl\{ \begin{bmatrix} \ket{T_x}  & \ket{T_y}& \ket{T_z} \\ \end{bmatrix}
 Z^\dagger R - \begin{bmatrix} \ket{S_x}  & \ket{S_y}& \ket{S_z} \\ \end{bmatrix}
Z^\dagger E R \biggr\}
\label{wckm33}
\end{eqnarray} 
where the matrices 
\begin{eqnarray} 
E:=\begin{bmatrix}
   \epsilon_1 & 0 & 0  \\
    0 & \epsilon_2 & 0  \\
    0 &  0 & \epsilon_3  \\
\end{bmatrix}
\qquad\qquad
R:=\begin{bmatrix}
   \frac{1}{\sqrt{1+\epsilon_1^2}} & 0 & 0  \\
    0 & \frac{1}{\sqrt{1+\epsilon_2^2}} & 0  \\
    0 &  0 & \frac{1}{\sqrt{1+\epsilon_3^2}} \\
\end{bmatrix}
\label{abktrua}
\end{eqnarray} 
have been introduced.

Just as in the case of leptons (\ref{dyop1}) there is a 1-dimensional dyadic transformation
\begin{eqnarray} 
\mathcal{O} = 
\begin{bmatrix} \ket{u_w}  & \ket{c_w}& \ket{t_w} \\ \end{bmatrix}
\begin{bmatrix} \bra{d_w}  \\ \bra{s_w} \\ \bra{b_w} \\ \end{bmatrix}
\label{dyop2}
\end{eqnarray} 
which operates between the 3-dimensional spaces of up- and down-type quark states. Due to quark-lepton universality, when expressed in terms of operators $\vec S$ and $\vec T$, the operator $\mathcal{O}$ for quarks must be identical to what was used for leptons in (\ref{wb1}).

Restricting, for a moment, on the dot and triple product contributions (\ref{wb1}) as input, one may then calculate $V_{CKM}$ given in (\ref{wckm33}) to be 
\begin{eqnarray} 
V_{CKM}=I + RZX Z^\dagger E R + REZXZ^\dagger R
\rightarrow \exp \{RZX Z^\dagger E R + REZXZ^\dagger R\}
\label{wck41}
\end{eqnarray} 
where I is the 3$\times$3 unit matrix arising from the dot product terms in (\ref{wb1}). The other terms in (\ref{wck41}) are the anti-hermitian contributions from the triple product in (\ref{pol12}) and (\ref{wb1}). They replace the expression $ZXZ^\dagger$ in (\ref{pmn00a}) for leptons. 
%To derive (\ref{wck41}) one should note that the $O(E^0)$ and $O(E^2)$ terms in (\ref{wckm33}) exactly combine to give the unit matrix in (\ref{wck41}). 

Just as in the case of leptons one may improve on this result by including the contributions from (\ref{w8899}), in order to obtain the desired non-symmetric contributions to $|V_{CKM}|$. The improved formula for the CKM matrix reads
\begin{eqnarray} 
V_{CKM}=\exp \{ 2[RZD Z^\dagger E R - REZD^\dagger Z^\dagger R] \}
\exp \{RZX Z^\dagger E R + REZXZ^\dagger R\}
\label{wck41111}
\end{eqnarray} 
In contrast to X in (\ref{matz325}) the matrix D in (\ref{ddma1}) is not anti-hermitian. This fact has been accounted for in the first exponential factor.

Eq. (\ref{wck41111}) allows to evaluate $|V_{CKM}|$ using appropriate values for the fermion masses entering (\ref{sm14exxx}). It must be noted, however, that the low energy values of the $\epsilon_{i}$ are not useful in this context. Instead one should use running masses near the Planck scale, because the dynamics generates fermion masses originally at Planck scale distances\footnote{A GUT scale is not present in the model. There is only the Fermi scale, defined as the interaction energy of the isospin vectors, and the Planck scale, defined as the binding energy of the fields $\Psi$\cite{bodohiggs}.}. Unfortunately, the predictions for running masses are not very precise because higher order contributions become appreciable at very large scales. Nevertheless, I am using results from the literature\cite{runnmass,juarez} to determine the $\epsilon_{i}$ at high scales.
\begin{eqnarray} 
\epsilon_{1} = 0.35  \qquad \epsilon_{2} =0.070 \qquad \epsilon_{3} =0.0040
\label{sm15}
\end{eqnarray}
unfortunately with a large theoretical error\cite{juarez}, whose magnitude even is hard to estimate. The numbers are for a 2HDM (2 Higgs doublet model) which is known to be the low-energy limit of the microscopic model\cite{bodohiggs}. They exhibit a family hierarchy which will be seen to induce a corresponding hierarchy in the mixing of the quark families. Actually, as discussed in earlier work\cite{bodomasses}, this is to be expected within the present model due to the large top mass which forces the up- and down-type mass eigenstates to be approximately $\sim\vec S$ and $\sim\vec T$, respectively, in (\ref{allup}) and (\ref{alldown}), much unlike the lepton states which are $\sim\vec S \pm \vec T$ according to (\ref{fo77}).
%(bessser mit meinen Zahlen aus dem Block eps2=0.06GUTbis0.09Planck und eps3=0.003GUTbis0.005Planck erlaubt. eps1=0.33 bedeudet mu+md=0.25MeV. In der einen running Arbeit gehen mu und md recht stark herunter, erst recht bis zur Planck scale, und dort kann higher orders noch stärker sein)

Just as masses, CKM matrix elements are running, i.e. dependent on the scale paramter $t=\ln \frac{E}{\mu}$ where E is the relevant energy scale and $\mu$ the renormalization scale. The running of the absolute values of the CKM matrix elements has been discussed for the 2HDM in \cite{juarez}. It turns out to be remarkably simple, at least in leading order, because it can be given in terms of one universal function h(t). 
\begin{eqnarray}
|V_{CKM}(t)| \approx  
\begin{bmatrix}
|V_{ud}(0)| & |V_{us}(0)| & \frac{|V_{ub}(0)|}{h(t)} \\
|V_{cd}(0)| & |V_{cs}(0)| & \frac{|V_{cb}(0)|}{h(t)} \\
\frac{|V_{td}(0)|}{h(t)} & \frac{|V_{ts}(0)|}{h(t)} & |V_{tb}|(0)
\end{bmatrix} 
\label{mmccrr1}
\end{eqnarray}
For the Jarlskog invariant one has
\begin{eqnarray}
J_{CKM}(t) \approx \frac{J_{CKM}(0)}{h^2(t)} 
\label{mmccrjr}
\end{eqnarray}
In the 2HDM case $h(t)$ is a moderately varying function. According to \cite{juarez} it increases by about 20\% when going from GeV to Planck scale energies.
%Numerical results on the function h(t) can be found, for example, in ***cite. Unfortunately, its behavior strongly depends on which kind of low energy model one prefers. h(t) is decreasing for the SM and increasing for the 2HDM. More precisely, h(t) decreases by almost a factor of 2 in the SM when going from GeV energies to Planck scale energies, while it increases only moderately by about 20\% in the 2HDM case.

Using (\ref{wck41111}) and (\ref{sm15}) I have calculated the CKM elements at high energies and then extrapolated them back to GeV energies according to (\ref{mmccrr1}). I obtain the matrix $|V_{CKM}|$ of absolute values 
\begin{eqnarray} 
\begin{bmatrix}
   0.974 & 0.224 & 0.0035  \\
    0.224 & 0.973 & 0.044  \\
      0.0080 &  0.043 & 0.9991  \\
\end{bmatrix}
%\quad vs. \quad
vs.
\begin{bmatrix}
   0.9734-0.9740 & 0.2235-0.2251 & 0.00362-0.00402  \\
    0.217-0.225 & 0.969-0.981 & 0.0394-0.0422  \\
      0.0083-0.0088 &  0.0404-0.0424 & 0.985-1.043  \\
\end{bmatrix}
\label{ckm1aa}
\end{eqnarray} 
The numbers look reasonable, as compared to the phenomenological values, and show the correct hierarchy and orders of magnitude. However, the theoretical uncertainty from the scale evolution is large and difficult to estimate, in particular concerning quark mass values near the Planck scale. For example, $\epsilon_{1}$ accommodates the Cabbibo angle correctly, whereas the `23'-matrix elements $|V_{ts}|$ and $|V_{cb}|$ tendencially come out too large, while the `13'-elements $|V_{ub}|$ and $|V_{td}|$ are typically too small. These deviations may seem being just 2$\sigma$ effects, but as stressed before the theoretical error from the quark mass evolution is extremely difficult to handle.
%However, such a large $\epsilon_{1}$ is obtained from my programs only, if I would bring down both the up- and the down-quark mass values to below 1 MeV, more precisely to $M_u+M_d\approx 0.5$ MeV, using (\ref{sm14ex}).
%It thus seems that for the light quarks the linear approximation (\ref{eqxxdrt1}) used throughout this work is not valid. Next-to-leading effects from the heavy families could modify severely the contributions from the light quark masses. Alternatively, one may consider the possibility that the so-called `current mass' values for up and down quark cannot be applied here. I am still working on the problem to find a satisfactory solution. 

Similarly, concerning the Jarlskog invariant one obtains $J_{CKM}=0.000027$, a bit small when compared to the observed value $J_{CKM}=(3.00+0.15-0.09) \times 10^{-5}$.%The analytic resutl obtained from can be brought to Wolfenstein form by multiplying the analytic result form (\ref{wck41111}) by suitable phases and defining the Wolfenstein parameters in terms of the epsi as follows.

\begin{figure}[t]
\centering
  \begin{tabular}{@{}ccc@{}}
    \includegraphics[width=.30\textwidth]{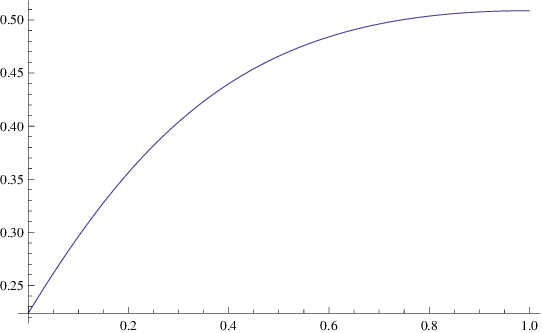} &
    \includegraphics[width=.30\textwidth]{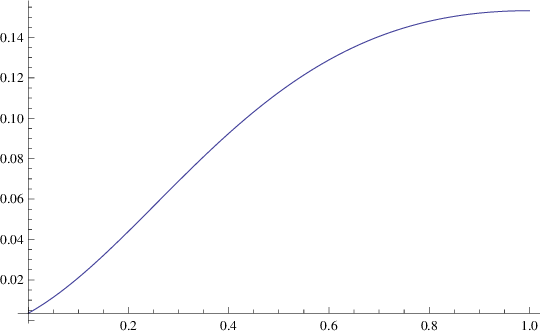} &
    \includegraphics[width=.30\textwidth]{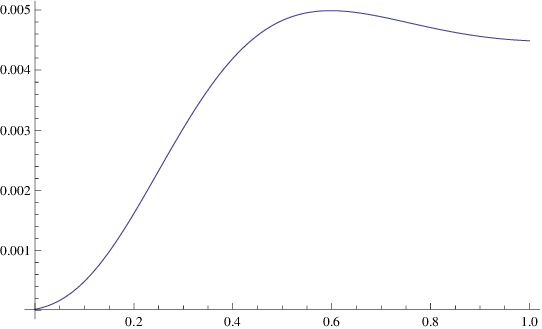}   
  \end{tabular}
\caption{Transition between the CKM and the PMNS limit of the matrix elements $|V_{12}|$ and $|V_{13}|$ and the Jarlskog invariant (from left to right) as a function of the parameter $\alpha$ defined in the main text. For example, $|V_{12}|$ starts with its CKM value 0.224 at $\alpha =0$ and grows towards the PMNS value at $\alpha =1$.}
\end{figure}

In conclusion, explicit analytic and numerical results for the mixing matrices have been presented in this work. Of particular interest are the prediction for the PMNS matrix (\ref{p0imp}) and the fermion mass dependence of the CKM matrix as given by (\ref{wck41111}). Actually, (\ref{wck41111}) is universal in that it embraces (i) the case of no mixing ($\epsilon_{1}=\epsilon_{2}=\epsilon_{3}=0$), (ii) the CKM prediction obtained with $\epsilon_{i}$-values (\ref{sm15}) and (iii) the PMNS formula which formally is given using $\epsilon_{1}=\epsilon_{2}=\epsilon_{3}=1$. To make this visible, I have drawn in Fig. 1 the `12' (i.e. Cabibbo) and the `13' matrix element and the Jarlskog invariant as a function of a parameter $\alpha$. $\alpha$ is introduced to avoid drawing the full $\epsilon_{i}$-dependence of the matrix elements and defined in such a way that it vanishes in the CKM case and takes the value of 1 in the PMNS limit. More precisely, one has 
\begin{eqnarray} 
\epsilon_{1} = 0.35 + 0.65\,\alpha   \qquad \epsilon_{2} =0.07 + 0.93\,\alpha  \qquad \epsilon_{3} =0.004 + 0.996\,\alpha 
\label{sal15}
\end{eqnarray}

%as well known the value of JCKM is not desribed correctly by the Wolfenstein matrix because it is order lambda4.

%\begin{figure}[h]
%\begin{center}
%\includegraphics[width=4.8in]{v12forletter}
%\end{center}
%\caption{***v12 and 23 with positive slope, 13 with negative. gleichartig zwischen aCKM=0 und aPMNS=1 extrapoliert: epp1 = 0.367 + a 0.633 epp2 = 0.05 + a 0.95 epp3 = 0.004 + a 0.996. ep1=ep2=ep3=0 ist kein Mixing v12=0 usw}
%\label{aba:fig1}
%\end{figure}

%-------------------------------------------------------------


\begin{thebibliography}{99}
\bibitem{maki} Z. Maki, M. Nakagawa and S. Sakata, Progr.Theor. Phys. 28 (1962) 870.
\bibitem{giganti} C. Giganti, S. Lavignac and M. Zito, Progr. Part. Nucl. Phys 98 (2018) 1.
\bibitem{jarls} C. Jarlskog, Z. Phys. C29 (1985) 491.
\bibitem{pdg} Particle Data Group, Prog. Theor. Exp. Phys. 8 (2022 and 2023 update) 083C01.
\bibitem{jarl2}  A. R. Ellis, K. J. Kelly and S. W. Li, Phys. Rev. D 102 (2020) 115027.
\bibitem{bodohiggs} B. Lampe, Progr. of Phys. 69 (2021) 2000072.
\bibitem{bodomasses} B. Lampe, Int. J. Mod. Phys. A31 (2016) 1650115, 1650116.
%\bibitem{bodotalk} B. Lampe in Proc. of the MG16 Meeting on General Relativity, doi.org/10.1142/13149 (2023) 999.
%\bibitem{bodoreview} see the review B. Lampe, arXiv:1505.03477 [hep-ph] (2015).
%\bibitem{bodotalk} B. Lampe in Proc. of the MG16 Meeting on General Relativity, 5–10 July 2021, editors R. Ruffini and G. Vereshchagin, https://doi.org/10.1142/13149 (2023) 999.
\bibitem{cab} N. Cabibbo, Phys. Rev. Lett. 10 (1963) 531.
\bibitem{runnmass} K. Bora, arXiv:1206.5909 [hep-ph] (2012).
\bibitem{juarez} W. Juarez et al., Phys. Rev. D66 (2002) 116007. 
\bibitem{bodoprep} B. Lampe, in preparation.
\end{thebibliography}
\end{document}